\documentclass[12pt,preprint]{aastex}  

\newcommand{\va}{v_{\mathrm{A}}}
\newcommand{\cs}{c_{\mathrm{s}}}
\newcommand{\ct}{c_{\mathrm{T}}}
\newcommand{\nct}{\tilde{c}_{\mathrm{T_0}}}

\newcommand{\nctp}{\tilde{c}_{\mathrm{T_p}}}
\newcommand{\nctc}{\tilde{c}_{\mathrm{T_c}}}

\newcommand{\vap}{v_{\mathrm{Ap}}}
\newcommand{\csp}{c_{\mathrm{sp}}}
\newcommand{\ctp}{c_{\mathrm{Tp}}}

\newcommand{\lambdamod}{\tilde{\Lambda}_0}

\newcommand{\lambdamodp}{\tilde{\Lambda}_{\rm p}}
\newcommand{\lambdamodc}{\tilde{\Lambda}_{\rm c}}

\newcommand{\vac}{v_{\mathrm{Ac}}}
\newcommand{\csco}{c_{\mathrm{sc}}}
\newcommand{\ctc}{c_{\mathrm{Tc}}}
\newcommand{\tildectc}{\tilde{c}_{\mathrm{Tc}}}

\newcommand{\der}{{\rm d}}
\newcommand{\pd}{\partial}
\newcommand{\td}{\tau_{\mathrm{D}}}
\newcommand{\ck}{c_{\rm k}}

\newcommand{\Omegao}{\Omega_0}

\begin{document}

	\title{Non-adiabatic magnetohydrodynamic waves in a cylindrical prominence thread with mass flow}

	\shorttitle{Non-adiabatic magnetohydrodynamic waves in a cylindrical  prominence thread with mass flow}

   \author{R. Soler, R. Oliver, and J. L. Ballester}

   \affil{Departament de F\'isica, Universitat de les Illes Balears,
              E-07122, Palma de Mallorca, Spain}
              \email{[roberto.soler;ramon.oliver;joseluis.ballester]@uib.es}

  \begin{abstract}
 High-resolution observations show that oscillations and waves in prominence
threads are common and that they are attenuated in a few periods. In addition, observers have also reported the presence of material flows in such prominence fine-structures. Here we investigate the time damping of non-leaky oscillations supported by a homogeneous cylindrical prominence thread embedded in an unbounded corona and with a steady mass flow. Thermal conduction and radiative losses are taken into account as damping mechanisms, and the effect of these non-ideal effects and the steady flow on the attenuation of oscillations is assessed. We solve the general dispersion relation for linear, non-adiabatic magnetoacoustic and thermal waves supported by the model, and find that slow and thermal modes are efficiently attenuated by non-adiabatic mechanisms. On the contrary, fast kink modes are much less affected and their damping times are much larger than those observed. The presence of flow has no effect on the damping of slow and thermal waves, whereas fast kink waves are more (less) attenuated when they propagate parallel (anti-parallel) to the flow direction. Although the presence of steady mass flows improves the efficiency of non-adiabatic mechanisms on the attenuation of transverse, kink oscillations for parallel propagation to the flow, its effect is still not enough to obtain damping times compatible with observations.
  \end{abstract}



\section{Introduction}

Prominences and filaments are large-scale magnetic structures embedded in the solar corona, and whose plasma density and temperature are
akin to those of the chromosphere. High-resolution images of solar filaments \citep[e.g.,][]{lin2003, lin2005, lin2007} clearly show the existence of
horizontal fine-structures within the filament body. This observational evidence suggests that prominences are
composed of many field-aligned threads. These threads are usually skewed with respect to the filament long axis by an angle of 20$\degr$ on average, although their orientation can vary significantly within the same prominence \citep{lin2004}. The observed thickness, $d$, and length, $l$, of threads are typically in the ranges $0.2\, {\rm arcsec} < d <  0.6\, {\rm arcsec}$ and $5\, {\rm arcsec} < l <  20\, {\rm arcsec}$ \citep{lin2005}. Since the observed thickness is close to the resolution of present-day telescopes, it is likely that even thinner threads could exist.  According to some models \citep[e.g.,][]{ballesterpriest}, a thread is believed to be part of a larger magnetic coronal flux tube which is anchored in the photosphere, with denser and cooler material near its apex, i.e. the observed thread itself. However, the process that leads to the formation of such structures is still unknown.

There are many evidences of small-amplitude waves and oscillatory motions in quiescent prominences \citep[this topic has been reviewed by][]{oliverballester02, ballester, banerjee}. Focusing on prominence fine-structures, some observers have detected oscillations and traveling waves in individual threads or groups of threads \citep{yi1,yi2,lin2004,lin2007}, with periods typically between 3 and 20~min. In addition, mass flows along filament threads, with a flow velocity in the range $5$~--~$25$~km~s$^{-1}$, have been also observed \citep{zirker,lin2003,lin2005}. Moreover, it is noticeable that \citet{zirker} and \citet{lin2003} have detected flows in opposite directions within adjacent threads, a phenomenon known as counter-streaming. On the other hand, some observational works have suggested signatures of wave damping in prominence oscillations \citep{landman,tsubaki,lin2004}, but to date only \citet{molowny} and \citet{terradasobs} have studied in detail this phenomenon, in particular in two-dimensional Doppler velocity time series. This analysis showed that oscillations detected in large areas of a quiescent prominence were attenuated after few periods. Although this quick attenuation seems to be a common feature of prominence oscillations, unfortunately no similar observational study focusing on the attenuation of individual thread oscillations has been performed yet.

From the theoretical point of view, the usual interpretation of thread oscillations is in terms of the adiabatic magnetohydrodynamic eigenmodes supported by the thread body. The first investigation of individual thread vibrations was performed by \citet{joarder}, whose work was extended and corrected by \citet{diaz2001}, considering a non-isothermal Cartesian thread surrounded by the coronal medium \citep[based on the model by][]{ballesterpriest}, in the $\beta=0$ approximation. These authors found that only the low-frequency oscillatory modes are confined within the dense region, and that perturbations can achieve large amplitudes in the corona at long distances from the thread. Later, \citet{diaz2003} assumed the same geometry, but took longitudinal propagation into account, and obtained a better confinement for the perturbations. Considering a more realistic and representative cylindrical geometry, \citet{diaz2002} found that a non-isothermal cylindrical thread supports an even smaller number of trapped oscillations and that perturbations are much more efficiently confined within the cylinder in comparison with the Cartesian case. It is worth to mention that the collective oscillations of multithread systems have also been investigated by \citet{diaz2005} and \citet{diazroberts}, again in Cartesian geometry. See \citet{ballester} for a review about theoretical works.

The effect of steady mass flows on the oscillatory modes of magnetic structures has been theoretically investigated in some works. The most relevant ones for the present investigation are \citet{naka}, who studied the effect of steady flow in coronal and photospheric slabs, and \citet{terra}, who extended the former study to cylindrical geometry. In addition to producing a shift of the oscillatory frequency, both papers show that the main effect of the flow is to break the symmetry between parallel and anti-parallel wave propagation to the flow direction and, for sufficiently strong flows, slow modes can only propagate parallel to the flow direction, anti-parallel propagation being forbidden.

Turning to the damping of oscillations, its theoretical investigation has been undertaken by a number of recent
papers. Among the proposed damping mechanisms to explain the attenuation \citep{ballai}, non-adiabatic effects are the most
extensively investigated to date, although other candidates like wave leakage \citep{schutgensA,schutgensB,schutgensToth} and ion-neutral collisions \citep{forteza} have been also studied. By removing the adiabatic assumption and taking into account thermal conduction and radiative losses as damping mechanisms, some works have studied the time damping in a homogeneous, unbounded plasma \citep{carbonell}, in an isolated prominence slab \citep{terradas}, and in a prominence slab embedded in the solar corona \citep{soler1,soler2}. The common main result of these studies is that non-adiabatic mechanisms can explain the observed damping times in the case of slow modes, whereas fast modes are much less attenuated by non-adiabatic effects. It is important to note that all these referred articles have studied the wave attenuation in models which attempt to represent the whole prominence body. Thus, none of them have neither considered the prominence fine-structure nor mass flows.

More recently, \citet{carbonellflow} have performed the first attempt to study the combined effect of both non-adiabatic mechanisms and steady flows on the time damping of slow and thermal waves in a homogeneous, unbounded prominence plasma. These authors found that the mass flow does not modify the damping time of both slow and thermal waves with respect to the case without flow, but the period of the slow wave increases dramatically for flow velocities close to the non-adiabatic sound speed. Moreover, the thermal disturbance behaves as a propagating mode in the presence of flow. The present work goes a step forward with respect to \citet{carbonellflow} since a more complicated geometry is assumed here. Our aim is to describe the effect of both mass flow and non-adiabatic effects on the oscillations supported by an individual prominence thread modeled as a homogeneous and infinite cylinder embedded in an unbounded and also homogeneous corona. For simplicity, the hot, coronal part of the magnetic tube that contains the thread is not taken into account. Gravity is also discarded. The inclusion of gravity would involve the consideration of a more complicated magnetic structure able to support the prominence material \citep[e.g.,][]{ballesterpriest, schmitt, rempel}. However, such kind of configurations are unstable and incomplete, since they do not incorporate the physics that lead to stable solutions and, moreover, the obtained prominence widths are much smaller than those observed. Although some prominences and threads show an unstable behavior, there is a number of observations of stable oscillating threads \citep[e.g.,][]{lin2003, lin2005, lin2007}. For this reason and since the present work is focused on the study of waves in stable threads, we neglect the effect of gravity and consider a stable simplified model that allows us to deal with wave solutions.

This paper is organized as follows. The description of the model configuration and the basic equations for the discussion of linear non-adiabatic waves are given in \S~\ref{sec:equilibrium}. Then, the results are presented in \S~\ref{sec:results}, first for the case without flow and later extended by including a steady mass flow in the equilibrium. Finally, \S~\ref{sec:conclusions} contains our conclusions.


\section{Model equations}
\label{sec:equilibrium}

\begin{figure}[!htb]
\centering
\epsscale{0.75}
\plotone{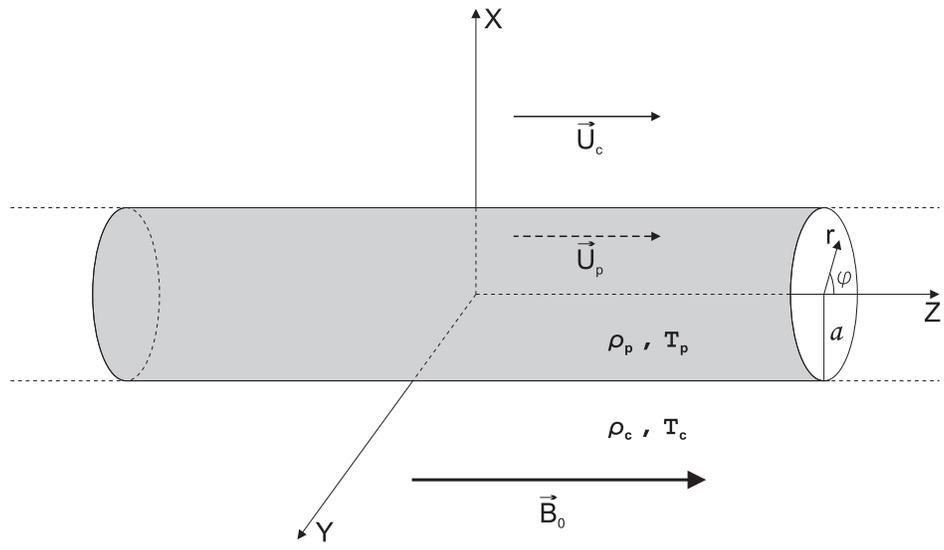}
\caption{Sketch of the model. \label{fig:model}}
\end{figure}

The model configuration considered in the present work (Fig.~\ref{fig:model}) is made of a homogeneous and isothermal plasma cylinder of radius $a$ with prominence conditions (density $\rho_{\rm p}$ and temperature $T_{\rm p}$) embedded in an unbounded corona (density $\rho_{\rm c}$ and temperature $T_{\rm c}$). The cylinder is also unlimited in the axial direction. The internal temperature and density, as well as the coronal temperature, are considered as free parameters. However, the value of the coronal density is given by imposing total pressure continuity across the interface between the flux tube and the external medium. In all the following expressions, a subscript  0 indicates local values, while subscripts p and c denote quantities explicitly computed with prominence and coronal parameters, respectively. 

\clearpage
\clearpage

Given the model geometry, we use cylindrical coordinates, namely $r$, $\varphi$, and $z$ for the radial, azimuthal and longitudinal coordinates, respectively. The magnetic field is uniform and orientated along the cylinder axis,  ${\mathit {\bf B}}_0=B_0 \hat{\bf e}_z$, $B_0$ being the same constant in the thread and in the coronal medium. A steady mass flow is assumed along the $z$-direction, whose flow velocity can be different in the cylinder and in the corona. Thus, ${\mathit {\bf U}}_{\rm p}=U_{\rm p} \hat{\bf e}_z$ and ${\mathit {\bf U}}_{\rm c}=U_{\rm c} \hat{\bf e}_z$ correspond to the steady flow in the flux tube and in the corona, respectively. 

Parallel thermal conduction to the magnetic field, radiative losses, and heating are considered as non-adiabatic effects. We assume that the plasma is fully ionized, and so the cross-field or perpendicular thermal conduction is absolutely negligible. The contribution of neutrals to the thermal conduction in a partially ionized plasma has been investigated by \citet[][in preparation]{fortezanonad}. Radiation and heating are evaluated together by means of the heat-loss function, $L(\rho,T)= \chi^* \rho T^\alpha - h \rho^{a^*} T^{b^*}$, in which radiation is parametrized with $\chi^*$ and $\alpha$ \citep{hildner} and the heating scenario is given by the exponents $a^*$ and $b^*$ \citep{rosner,dahlburg}. Since the results for different heating scenarios do not show significant differences for prominence conditions \citep{carbonell,terradas,soler1,soler2}, here we restrict ourselves to a constant heating per unit volume ($a^*=b^*=0$). 

The basic equations for the discussion of non-adiabatic oscillations supported by this equilibrium correspond to equations~(1)--(6) of \citet[hereafter Paper~I]{soler1}, whose linearized version applied to the present case is given next,
\begin{equation}
 \frac{\pd \rho_1}{\pd t} + U_0 \frac{\pd \rho_1}{\pd z} + \rho_0 \nabla \cdot {\mathit {\bf v}}_1 = 0,
\end{equation}
\begin{equation}
 \rho_0 \frac{\pd {\mathit {\bf v}}_1}{\pd t} + \rho_0 U_0 \frac{\pd {\mathit {\bf v}}_1}{\pd z} = - \nabla p_1 + \frac{1}{\mu} \left( \nabla \times {\mathit {\bf B}}_1 \right) \times {\mathit {\bf B}}_0,
\end{equation}
\begin{eqnarray}
 \frac{\pd p_1}{\pd t} &+& \rho_0 U_0 \frac{\pd p_1}{\pd z} - \cs^2 \left( \frac{\pd \rho_1}{\pd t} + \rho_0 U_0 \frac{\pd \rho_1}{\pd z} \right) \nonumber \\
 &+& \left( \gamma - 1 \right) \left[ \rho_0 L_\rho \rho_1 + \rho_0 L_T T_1 - \kappa_\parallel \frac{\pd^2 T_1}{\pd z^2}  \right] = 0,  \label{eq:energy}
\end{eqnarray}
\begin{equation}
 \frac{\pd {\mathit {\bf B}}_1}{\pd t} + U_0 \frac{\pd {\mathit {\bf B}}_1}{\pd z} = \nabla \times \left( {\mathit {\bf v}}_1 \times {\mathit {\bf B}}_0 \right),
\end{equation}
\begin{equation}
 \nabla \cdot {\mathit {\bf B}}_1 = 0,
\end{equation}
\begin{equation}
 \frac{p_1}{p_0} - \frac{\rho_1}{\rho_0} - \frac{T_1}{T_0} = 0. \label{eq:state}
\end{equation}
Here $p_0$, $\rho_0$, $T_0$, ${\mathit {\bf B}}_0$, and $U_0$ are the equilibrium gas pressure, density, temperature, magnetic field, and flow velocity, respectively. On the other hand, $p_1$, $\rho_1$, $T_1$, ${\mathit {\bf B}}_1 = B_r \hat{\bf e}_r + B_\varphi \hat{\bf e}_\varphi + B_z \hat{\bf e}_z$, and ${\mathit {\bf v}}_1 = v_r \hat{\bf e}_r + v_\varphi \hat{\bf e}_\varphi + v_z \hat{\bf e}_z$ are the linear gas pressure, density, temperature, magnetic field, and velocity perturbations, respectively. Finally, $\cs^2 = \frac{\gamma p_0}{\rho_0}$ is the adiabatic sound speed squared, $\kappa_\parallel  = 10^{-11} T^{5/2}\, \mathrm{W}\, \mathrm{m}^{-1}\, \mathrm{K}^{-1}$ is the thermal conductivity parallel to the magnetic field, and  $L_\rho$ and $L_T$ are the partial derivatives of the heat-loss function with respect to density and temperature, respectively (see Paper~I for details).

Assuming perturbations of the form $\tilde{f}_1 ( r, \varphi ) \exp \left( i \omega t - i k_z z \right)$, where $\omega$ is the (complex) oscillatory frequency and $k_z$ the (real) longitudinal wavenumber, one can combine equations~(\ref{eq:energy}) and (\ref{eq:state}) to obtain the following relation between the perturbed pressure and density,
\begin{equation}
 p_1 = \tilde{\Lambda}^2_0\, \rho_1, \label{eq:relprho}
\end{equation}
with
\begin{equation}
 \tilde{\Lambda}^2_0 \equiv \frac{\cs^2}{\gamma} \left[ \frac{\left( \gamma-1 \right) \left( \frac{T_0}{p_0} \kappa_\parallel k_z^2 + \omega_T - \omega_\rho \right) + i \gamma \Omega_0}
{\left( \gamma -1 \right) \left( \frac{T_0}{p_0} \kappa_\parallel k_z^2 + \omega_T \right) + i \Omega_0} \right], \label{eq:lambda}
\end{equation}
where $\Omega_0 = \omega - U_0 k_z$ is the Doppler-shifted frequency \citep{terra}, and $\omega_\rho$ and $\omega_T$ are defined in Paper~I. We see than the complex quantity  $\tilde{\Lambda}_0$ is a generalization, due to the presence of the flow, of the non-adiabatic sound speed defined in Paper~I. The real part of $\lambdamod$ plays the role of the sound speed when non-adiabatic effects are present. By means of this definition, one can see that the effect of non-adiabatic terms is to modify the medium sound speed, and so they most probably affect slow modes since they are mainly governed by acoustic effects. See \S~\ref{sec:soundspeed} for more details about this quantity.

Now, following \citet{terra}, we combine the basic equations and arrive to the following expressions,
\begin{equation}
\Upsilon^2 \left[ \Upsilon^2 - \left( \tilde{\Lambda}_0^2 + \va^2 \right) \nabla^2 \right] \Delta + \tilde{\Lambda}_0^2 \va^2 \frac{\pd^2}{\pd z^2} \nabla^2 \Delta = 0, \label{eq:basic}
\end{equation}
\begin{equation}
  \left( \Upsilon^2 - \va^2 \frac{\pd^2}{\pd z^2} \right) \Gamma = 0, \label{eq:alfven}
\end{equation}
where $\va^2 = \frac{B_0^2}{\mu \rho_0}$ is the Alfv\'en speed squared, $\Upsilon$ is a linear operator defined as follows,
\begin{equation}
 \Upsilon = \frac{\pd}{\pd t} + U_0 \frac{\pd}{\pd z},
\end{equation}
and $\Delta$ and $\Gamma$ are the divergence and the $z$-component of the rotational of the velocity perturbation, respectively,
\begin{equation}
 \Delta = \nabla \cdot {\mathit {\bf v}}_1,
\end{equation}
\begin{equation}
 \Gamma = \left( \nabla \times {\mathit {\bf v}}_1 \right) \cdot \hat{\bf e}_z.
\end{equation}
Equation~(\ref{eq:alfven}) is the same as equation~(14) of \citet{terra} and governs torsional, Alfv\'en waves, which are not damped by non-adiabatic mechanisms and so are not considered in the present investigation. On the other hand, equation~(\ref{eq:basic}) represents fast and slow magnetosonic waves, together with the thermal or condensation mode \citep{field}. If non-adiabatic terms are neglected, $\tilde{\Lambda}_0 = \cs$ and then our equation~(\ref{eq:basic}) reduces to equation~(13) of \citet{terra}, which in the absence of flow ($U_0 = 0$) is equivalent to the well-known equation~(17) of \citet{lighthill}.

Next, the cylindrical symmetry of the model allows us to write the divergence of the velocity perturbation in the following form,
\begin{equation}
 \Delta = R(r) \exp \left( i \omega t + i n \varphi - i k_z z \right), \label{eq:div}
\end{equation}
where $n$ is an integer that plays the role of the azimuthal wavenumber. Expressions for the perturbed quantities as a function of $\Delta$ can be found in Appendix~\ref{app:pert}. Now, applying this last expression to equation~(\ref{eq:basic}), one finds that $R(r)$ satisfies the well-known Bessel equation of order $n$,
\begin{equation}
 r^2 \frac{\der^2 R}{\der r^2} + r \frac{\der R}{\der r} + \left(m_0^2 r^2 - n^2 \right) R = 0, \label{eq:bessel}
\end{equation}
with
\begin{equation}
 m_0^2 = \frac{\left( \Omega_0^2 - k_z^2 \va^2 \right) \left( \Omega_0^2 - k_z^2 \tilde{\Lambda}^2_0 \right)}
{\left( \va^2 + \tilde{\Lambda}^2_0 \right) \left( \Omega_0^2 - k_z^2 \nct^2 \right)}, \label{eq:m0}
\end{equation}
\begin{equation}
\nct^2 \equiv \frac{\va^2 \tilde{\Lambda}^2_0}{\va^2 + \tilde{\Lambda}_0^2}. \label{eq:ct}
\end{equation}
The radial wavenumber squared, $m_0^2$, is in general a complex quantity, hence no pure body-like or surface-like waves are possible in non-adiabatic magnetohydrodynamics. If one assumes that $| \Re(m_0^2) | > | \Im(m_0^2) |$ (i.e., non-adiabatic effects produce a small correction to the adiabatic wave modes), the dominant wave character depends on the sign of $\Re(m_0^2)$. Thus, oscillations are {\em mainly} body-like if $\Re(m_0^2) > 0$ and solutions of  equation~(\ref{eq:bessel}) are Bessel functions. On the contrary, if $\Re(m_0^2) < 0$  oscillations are {\em mainly} surface-like (or evanescent) and solutions of  equation~(\ref{eq:bessel}) are modified Bessel functions. In this work we assume no wave propagation in the coronal medium, so the evanescent condition in the corona is imposed on the perturbations, namely $\Re(m_{\rm c}^2) < 0$. On the other hand, the present ordering of sound and Alfv\'en speeds does not permit the existence of surface waves within the fibril, so $\Re(m_{\rm p}^2) > 0$ is assumed. Then $R(r)$ is a piecewise function,
\begin{equation}
 R(r) = \left\{ \begin{array}{lcl} 
                 A_1 J_n (m_{\rm p} r ) & {\rm if} & r \le a, \\
		 A_2 K_n (n_{\rm c} r ) & {\rm if} & r > a,
                \end{array} \right.
\end{equation}
with $n_{\rm c}^2 = - m_{\rm c}^2$, $A_1$ and $A_2$ being complex constants. $J_n$ and $K_n$ are the usual Bessel and modified Bessel functions of order $n$, respectively \citep{abramowitz}. In order to obtain the dispersion relation that governs the behavior of wave modes, we impose the continuity of the Lagrangian radial displacement, $v_r / \Omega_0$, and the total pressure perturbation, $p_{\rm T_1}$, at the cylinder edge, $r = a$. After some algebra, the following expression is obtained,
\begin{equation} 
\frac{\rho_{\rm c}}{\rho_{\rm p}} \left( \Omega_{\rm c}^2 - k_z^2 \vac^2 \right) m_{\rm p} \frac{J'_n \left( m_{\rm p} a \right)}{J_n \left( m_{\rm p} a \right)} = \left( \Omega_{\rm p}^2 - k_z^2 \vap^2 \right) n_{\rm c} \frac{K'_n \left( n_{\rm c} a \right)}{K_n \left( n_{\rm c} a \right)}, \label{eq:disper}
\end{equation}
where $' \equiv \der /\der r$. Equation~(\ref{eq:disper}) is formally identical to equation~(21) of \citet{terra} since our non-adiabatic terms are enclosed in the definition of $m_{\rm p}$ and $n_{\rm c}$. If both non-adiabatic effects and the flow are dropped, equation~(\ref{eq:disper}) simply reduces to the well-known dispersion relation of \citet{edwinroberts}.

The solution of  equation~(\ref{eq:disper}) for a real $k_z$ is a complex frequency, $\omega = \omega_{\rm R} + i \omega_{\rm I}$. The oscillatory period ($P$), damping time ($\td$), and the ratio of both quantities are computed as follows,
\begin{equation}
P = \frac{2 \pi}{| \omega_{\rm R} |}, \qquad \tau_{\rm D} = \frac{1}{\omega_{\rm I}}, \qquad \frac{\tau_{\rm D}}{P} = \frac{1}{2\pi} \frac{| \omega_{\rm R } |}{\omega_{\rm I}}.
\end{equation}


\section{Results}
\label{sec:results}

\subsection{Configuration without flow}

In this section, we first perform a study of the solutions of the dispersion relation (eq.~[\ref{eq:disper}]) in the absence of steady flow. In this case, $U_{\rm p} = U_{\rm c} = 0$ and so $\Omega_{\rm p} = \Omega_{\rm c} = \omega$. This situation corresponds to the case studied by \citet{edwinroberts} with the addition of non-adiabatic effects. Unless otherwise stated,  the following equilibrium parameters are considered in all computations: $T_{\rm p}=8000$~K, \mbox{$\rho_{\rm p} = 5 \times 10^{-11}$~kg~m$^{-3}$}, $T_{\rm c} = 10^6$~K, $\rho_{\rm c} = 2.5 \times 10^{-13}$~kg~m$^{-3}$, $B_0=5$~G, and $a = 30$~km. Thus, the characteristic speeds of the internal and external media are: $\ctp = 11.56\, \mathrm{km}\, \mathrm{s}^{-1}$, $\csp = 11.76\,\mathrm{km}\, \mathrm{s}^{-1}$, $\vap =63.08\,\mathrm{km}\, \mathrm{s}^{-1}$, $\ctc = 163.51\, \mathrm{km}\, \mathrm{s}^{-1}$, $\csco =166.33\,\mathrm{km}\, \mathrm{s}^{-1}$, and  $\vac =892.06\,\mathrm{km}\, \mathrm{s}^{-1}$. In addition, both prominence and coronal plasmas are taken as optically thin (see Table~I in Paper~I for the values of parameters $\chi^*$ and $\alpha$ of the heat-loss function).

In the absence of flow, the complex oscillatory frequencies obtained by solving equation~(\ref{eq:disper}) for a fixed, real, and positive $k_z$ appear in pairs, $\omega_1 = \omega_{\rm R} + i \omega_{\rm I}$ and $\omega_2 = -\omega_{\rm R} + i \omega_{\rm I}$. The solution $\omega_1$ corresponds to a wave propagating towards the positive $z$-direction (parallel to magnetic field lines), whereas $\omega_2$ corresponds to a wave that propagates towards the negative $z$-direction (anti-parallel to magnetic field lines). For short, we call them parallel and anti-parallel waves, respectively. Both parallel and anti-parallel wave modes are equivalent and show exactly the same physical properties when no flow is considered. For the sake of simplicity, the results presented in this section correspond to parallel waves, which have a positive phase speed. Equivalent results for anti-parallel waves are deduced by considering negative phase speeds.

\subsubsection{Dispersion diagram and eigenfunctions of magnetoacoustic modes}

Magnetoacoustic modes supported by a magnetic cylinder have been extensively investigated \citep[e.g.,][]{spruit,edwinroberts,cally}. Fast oscillations with $n=0$, $n=1$, and $n \ge 2$ correspond to sausage, kink, and fluting or ballooning modes, respectively. On the basis that thread oscillations are observed in Doppler time series, this work is mainly focused on kink modes, which produce displacements of the thread axis from its original position. The fundamental fast kink mode is trapped for realistic values of the thickness and width of threads. Regarding slow modes, the fundamental modes and their harmonics are all trapped for any value of $k_z$ and $n$, but all of them have an almost identical frequency, as in the slab case (Paper~I). Thus, we also restrict ourselves to the fundamental slow mode with $n=1$ for simplicity. An additional solution of equation~(\ref{eq:disper}) is the thermal mode which, in the absence of flow, has a purely imaginary frequency (see \S~\ref{sec:thermal}).

\begin{figure}[!htb]
\centering
\epsscale{0.75}
\plotone{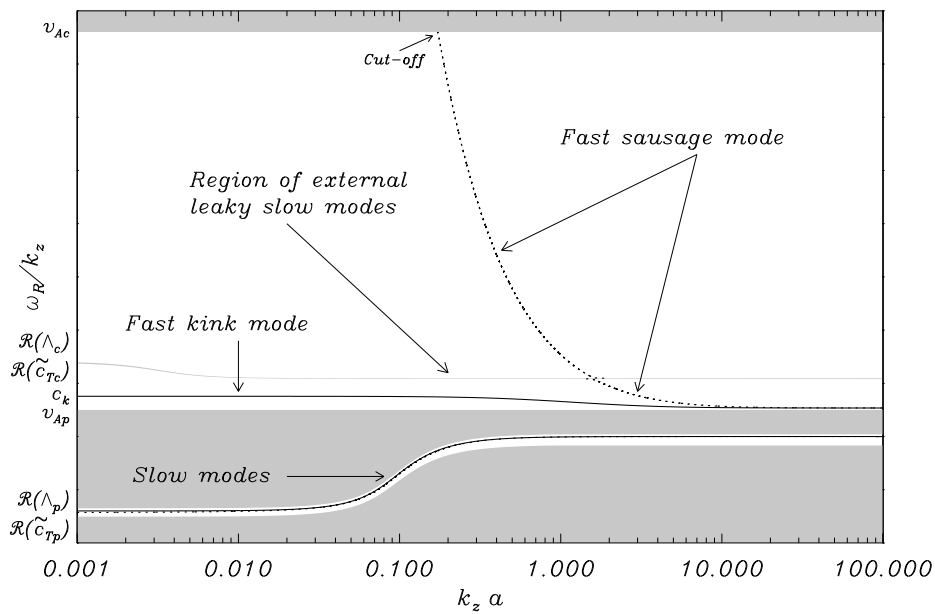}
\caption{Phase speed ($\omega_{\rm R}/k_z$) versus the dimensionless longitudinal wavenumber ($k_z a$) for the fundamental magnetoacoustic modes with $n=1$ (solid line) and $n=0$ (dashed line). The shaded zones are projections of the leaky regions on the plane of this diagram. Note the cut-off frequency of the fundamental fast sausage mode. The vertical axis is not drawn to scale. \label{fig:dispersion}}
\end{figure}

Figure~\ref{fig:dispersion} displays the phase speed diagram corresponding to the fundamental modes with $n=0$ and $n=1$ (compare this with Fig.~2 of Paper~I). One can see that the behavior of slow modes and the fast sausage mode is similar to that in a slab. An important difference with the results of Paper~I is that in the cylindrical case the fast kink mode does not couple with the external leaky slow modes enclosed in the region $\Re ( \nctc ) < \omega_{\rm R}/k_z < \Re ( \tilde{\Lambda}_{\rm c} )$, because its phase speed in the long-wavelength limit is $\ck < \Re \left( \nctc \right)$, with $\ck^2 = \frac{\left( \rho_{\rm p} \vap^2 + \rho_{\rm c} \vac^2 \right)}{ \left(\rho_{\rm p} + \rho_{\rm c}\right)}$.

\begin{figure*}[!htb]
\centering
\epsscale{1}
\plotone{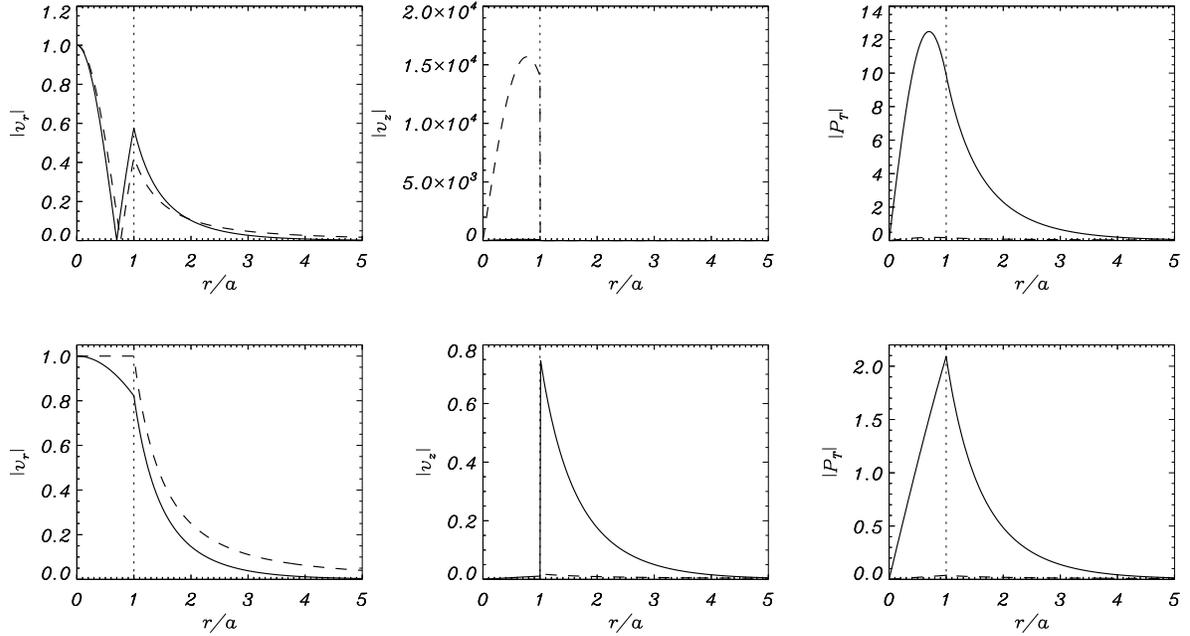}
\caption{Modulus of the eigenfunctions (in arbitrary units) corresponding to the radial velocity perturbation, $v_r$, the longitudinal velocity perturbation, $v_z$, and the total pressure perturbation, $P_{\rm T_1}$, as functions of the dimensionless distance to the cylinder axis for the fundamental slow (top panels) and fast kink (bottom panels) modes. The solid line corresponds to $k_z a=1$ while the dashed line corresponds to $k_z a = 10^{-2}$. Note that $v_z$ is not continuous at the cylinder edge, whose location is denoted by a vertical dotted line. \label{fig:autofun}}
\end{figure*}

Next, we plot in Figure~\ref{fig:autofun} the eigenfunctions corresponding to the radial and longitudinal velocity perturbations, $v_r$ and $v_z$, and the total pressure perturbation, $P_{\rm T_1}$, for the fundamental slow and fast kink magnetoacoustic modes. Contrary to the slab case of \citet[Fig.~4]{solerSF}, perturbations are efficiently confined within the cylinder for any value of the longitudinal wavenumber. Thus, this suggests that the influence of the corona on the damping of oscillations could be of smaller importance than in a magnetic slab. On the other hand, the expected velocity polarization is obtained, the slow mode being mainly polarized along the longitudinal direction ($v_z \gg v_r$) and the fast kink mode being responsible for transverse, radial motions ($v_r \gg v_z$).

\clearpage

\clearpage

\subsubsection{Damping times of fast kink and slow waves and comparison with a longitudinal slab}

\begin{figure*}[!htp]
\centering
\epsscale{1}
\plotone{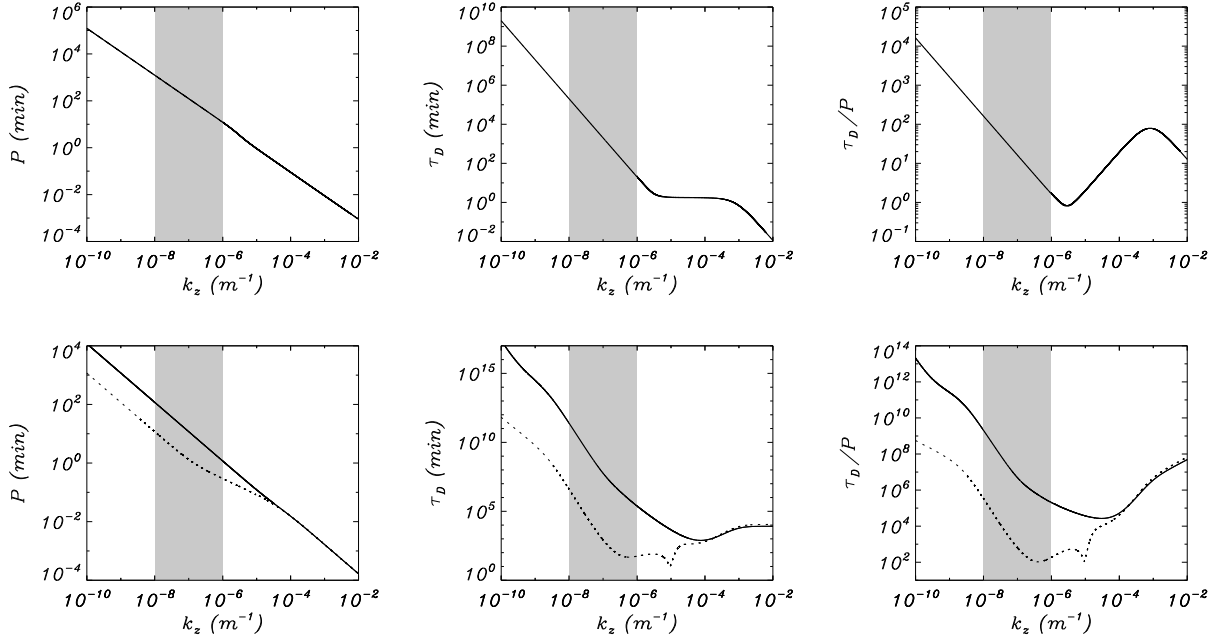}
\caption{Period (left), damping time (center), and ratio of the damping time to the period (right) versus the longitudinal wavenumber for the fundamental oscillatory modes in the absence of flow. Upper and lower panels correspond to the slow and fast kink modes, respectively. Solid lines are the solutions of the present, cylindrical equilibrium whereas dotted lines are the results of a longitudinal slab (Paper~I). Both curves overlap in the upper panels. Shaded zones indicate the range of observed wavelengths. \label{fig:nonadgen}}
\end{figure*}

We now compute the period, damping time, and their ratio for the fundamental fast kink and slow modes for a wide range of $k_z$ ($10^{-10}\, {\rm m}^{-1} < k_z < 10^{-2}\, {\rm m}^{-1} $). This range includes the values  corresponding to the observed wavelengths in prominences ($10^{-8}\, {\rm m}^{-1} \lesssim k_z \lesssim 10^{-6}\, {\rm m}^{-1} $). In Figure~\ref{fig:nonadgen} these results are compared with those obtained for a longitudinal slab whose half-width is equal to the cylinder radius (i.e., the case analyzed in \S~4.4 of Paper~I). We see that the results of the slow mode are almost identical in both cylindrical and slab geometries, hence the reader is referred to Paper~I for a description of the slow mode behavior. 

\clearpage
\clearpage

On the contrary, the results of the fast kink mode show significant differences between the slab and the cylinder. For small and intermediate $k_z$, both the period and damping time in a cylinder are larger than in a slab. This effect is specially noticeable for the damping time in the observed range of wavelengths, since its value in the cylindrical case is more than three orders of magnitude larger than in the slab. This causes the ratio $\td / P$ to be in the range $10^5 \lesssim \td / P \lesssim 10^9$ for the observed  wavelengths. Therefore, non-adiabatic effects are much less efficient in damping the fast kink mode in a cylinder than in the slab geometry. In addition, we see that the fast kink mode damping time due to non-adiabatic effects is much larger than typical lifetimes of filament threads and prominences. Hence, non-adiabatic fast kink waves are in practice undamped when these results are applied in the solar context.

\clearpage

\subsubsection{Thermal mode}
\label{sec:thermal}

Now we turn our attention to the thermal or condensation mode. The condensation instability has been studied in uniform, unbounded plasmas \citep{field}, in coronal slabs \citep{vanderlinden}, and in coronal cylinders \citep{an}. Since the thermal mode has a purely imaginary frequency, we now assume $\omega = i s$, where $s$ is real and often called the damping (or growing) rate. The situation $s > 0$ corresponds to a damped thermal mode, whereas $s < 0$ occurs if the mode is thermally unstable. The sign of $s$ can be estimated {\em a priori} by considering the stability criterion provided by \citet{field}, 
\begin{equation}
\kappa_{\parallel {\rm p}} k_z^2 + \rho_{\rm p} \left(  L_{T \rm{p}} - \frac{\rho_{\rm p}}{T_{\rm p}} L_{\rho \rm{p}} \right) > 0.
\end{equation}
Since for prominence conditions this inequality is verified for any real value of $k_z$, the thermal mode is always a damped solution and therefore we expect $s>0$. According to \citet{vanderlinden}, see Appendix~\ref{app:thermal} for more details, the evanescent assumption in the corona ($m_{\rm c}^2 < 0$) together with the body-wave assumption within the fibril ($m_{\rm p}^2 > 0$) is satisfied in a narrow range of $s$. However, there is an extremely narrow range of $k_z$ ($1.68 \times 10^{-7}\, {\rm m}^{-1} \lesssim k_z \lesssim 1.70 \times 10^{-7}\, {\rm m}^{-1} $) in which $m_{\rm p}^2 > 0$ and $m_{\rm c}^2 > 0$, and so the evanescent assumption is not verified. Then, the thermal mode does not exist as a non-leaky solution in such a range of $k_z$. Despite this, the fundamental thermal mode and all its harmonics have an almost identical damping rate $s$, whose value is also almost independent of the azimuthal wavenumber, $n$. For this reason and for the sake of simplicity, we again restrict ourselves to solutions with $n=1$ and focus on the fundamental mode.

\begin{figure*}[!htb]
\centering
\epsscale{1}
\plotone{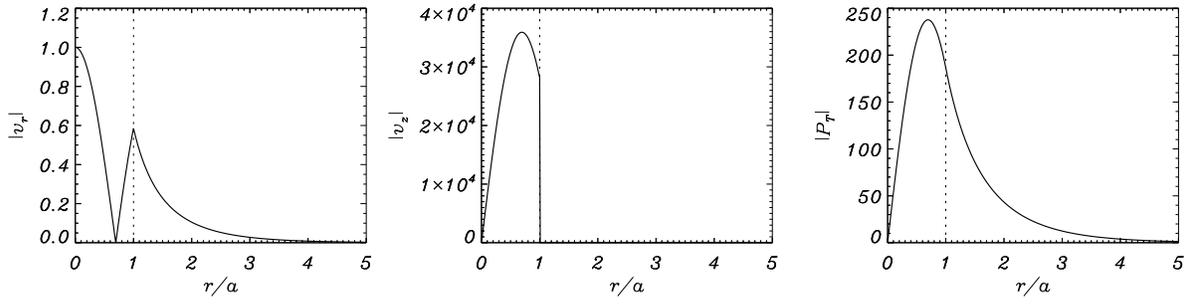}
\caption{Same as Fig.~\ref{fig:autofun} but for the fundamental thermal mode with $k_z a = 1$. \label{fig:autofunthermal}}
\end{figure*}

\begin{figure}[!tb]
\centering
\epsscale{0.5}
\plotone{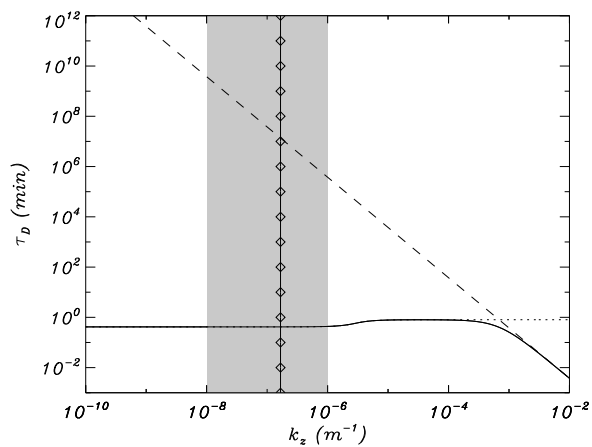}
\caption{Damping time versus the longitudinal wavenumber for the fundamental thermal mode. Different linestyles represent the omitted non-adiabatic mechanism: all mechanisms considered (solid line), prominence conduction eliminated (dotted line), prominence radiation eliminated (dashed line), coronal conduction eliminated (dot-dashed line), and coronal radiation eliminated (three dot-dashed line). The vertical line with symbols inside the shaded zone indicates the range of $k_z$ in which the thermal mode does not exist as a non-leaky solution (see Appendix~\ref{app:thermal} for details). \label{fig:thermal}}
\end{figure}

Figure~\ref{fig:autofunthermal} displays the spatial distribution of perturbations $v_r$, $v_z$, and $P_{\rm T_1}$ corresponding to the fundamental thermal mode with $n=1$. We see that the velocity field is dramatically polarized along the cylinder axis and that the eigenfunctions are very similar to those obtained for the slow mode (compare with Fig.~\ref{fig:autofun}, upper panels). On the other hand, the damping time ($\td = 1/s$) is plotted in Figure~\ref{fig:thermal} as a function of $k_z$. One can see that this mode is very quickly attenuated and that radiative losses from the prominence plasma are responsible for the attenuation in the observed wavelength range, whereas prominence thermal conduction is only relevant for large $k_z$. Coronal mechanisms have a negligible effect.

\clearpage
\clearpage

\clearpage
\clearpage


\subsection{Effect of steady flow}

Hereafter, we include a steady mass flow in the model in order to assess its influence on the previously described oscillatory modes. With no loss of generality, we assume no flow in the external medium, i.e. $U_{\rm c} = 0$. Moreover, the internal flow is assumed flowing towards the positive $z$-direction, i.e. $U_{\rm p} > 0$. Note that any other possible configuration can be obtained by means of a suitable election of the reference frame. 

\subsubsection{Phase speed shift}
\label{sec:phaseflow}

The reader is referred to \citet{terra} for a detailed description of the modification of the phase speed diagram due to the presence of flow. In short, the symmetry between parallel ($\omega_{\rm R} > 0$) and anti-parallel ($\omega_{\rm R} < 0$) waves is broken by the flow. Phase speeds of slow and fast parallel waves are now in the ranges $\left[ \Re \left( \nctp \right) + U_{\rm p} , \Re \left( \lambdamodp \right) + U_{\rm p} \right]$  and $\left[ \vap + U_{\rm p} , \vac \right]$, respectively. On the other hand, phase speeds of slow and fast anti-parallel waves lie now within the regions $\left[ - \Re \left( \lambdamodp \right) + U_{\rm p} , - \Re \left( \nctp \right) + U_{\rm p}  \right]$  and $\left[ -\vac , - \vap + U_{\rm p} \right]$, respectively. For a flow velocity larger than the internal non-adiabatic sound speed \citep[see][]{carbonellflow} the phase speed of anti-parallel slow waves is dragged to positive values, and so they become parallel waves in practice. These solutions were called backward waves by \citet{naka}. An equivalent phenomenon can also occur for fast waves for a superalfv\'enic flow, i.e. $U_{\rm p} > \vap$. Note that superalfv\'enic flows seem to be unrealistic in the light of observations.

Regarding thermal modes, the real part of their frequency now acquires a positive value, and their phase speed is equal to the flow velocity. Thus, thermal modes behave as propagating parallel waves with respect to the static, external reference frame. Although this result could be relevant for the observational point of view, since thermal modes might be detected as propagating waves in filament threads \citep{lin2007}, their extremely quick attenuation makes them undetectable in practice.

\clearpage
\begin{figure*}[!htp]
\centering
\epsscale{1}
\plotone{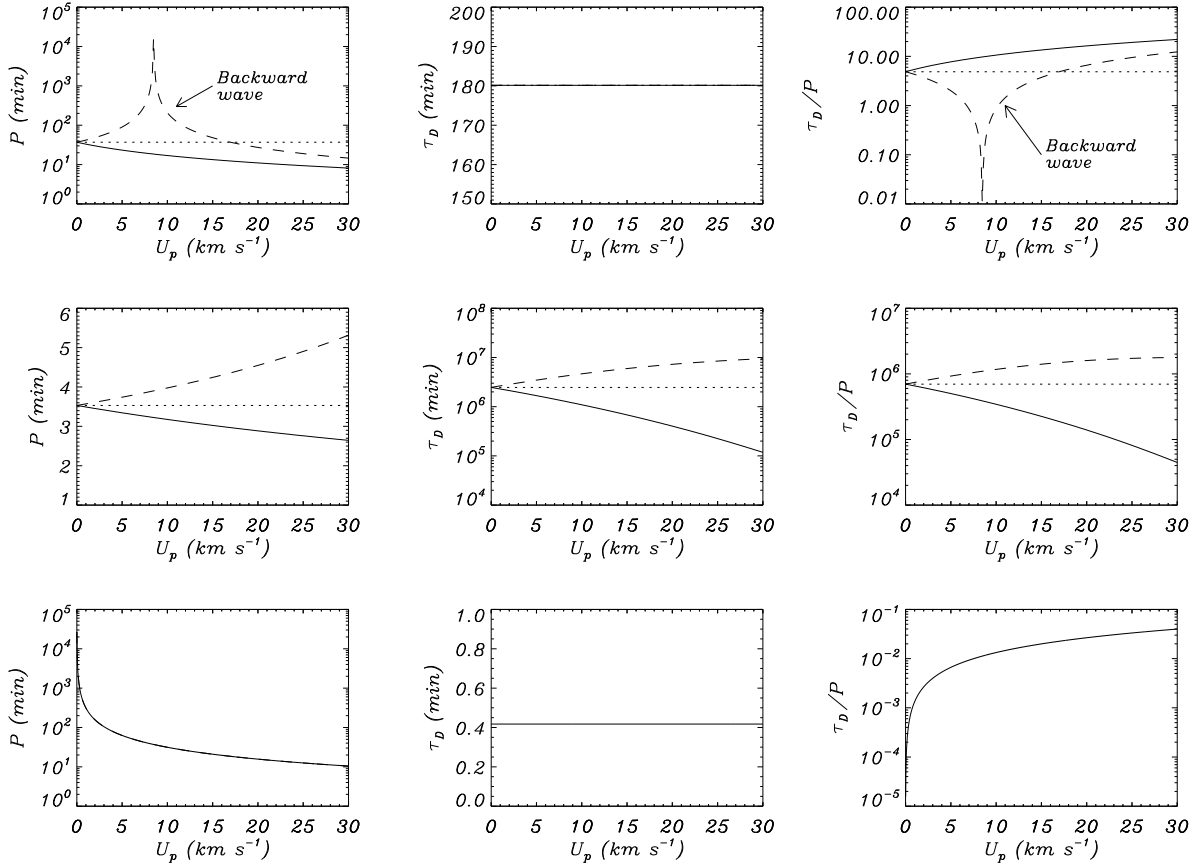}
\caption{Period (left), damping time (center), and ratio of the damping time to the period (right) versus the flow velocity for the fundamental oscillatory modes with $k_z a = 10^{-2}$. Upper, mid, and lower panels correspond to the slow, fast kink, and thermal modes, respectively. Different linestyles represent: parallel waves (solid line), anti-parallel waves (dashed line), and solutions in the absence of flow (dotted line). \label{fig:flowvar}}
\end{figure*}
\clearpage

\subsubsection{Influence on the damping time}

Regarding the effect of the flow on the damping time of oscillations, we plot in Figure~\ref{fig:flowvar} the dependence of the period, damping time, and their ratio as a function of the flow velocity. The longitudinal wavenumber has been fixed to $k_z a = 10^{-2}$, which corresponds to a value within the observed range of wavelengths. The flow velocity is considered in the range $0\, {\rm km}\, {\rm s}^{-1} < U_{\rm p} < 30\, {\rm km}\, {\rm s}^{-1}$, that corresponds to the observed flow speeds in quiescent prominences. In agreement with \citet{carbonellflow}, the anti-parallel slow wave becomes a backward wave for $U_{\rm p}  \approx 8.5$~km~s$^{-1}$, which corresponds to the non-adiabatic sound speed (eq.~[\ref{eq:lambda}]). This causes the period of this solution to grow dramatically near such flow velocity. However, the period of both parallel and anti-parallel fast kink waves is only slightly modified with respect to the solution in the absence of flow, and the thermal wave now has a finite period, which is comparable to that of the parallel slow mode.

\begin{figure*}[!htp]
\centering
\epsscale{1}
\plotone{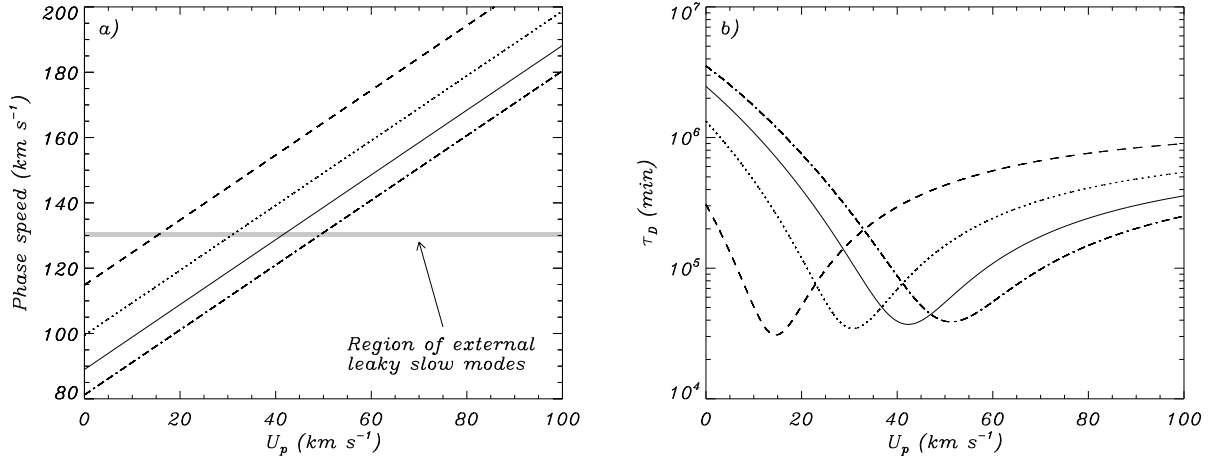}
\caption{$a)$ Phase speed and $b)$ damping time versus the flow velocity for the parallel fast kink mode with $k_z a = 10^{-2}$. Different linestyles represent different values of the prominence density: $\rho_{\rm p} = 5 \times 10^{-11}$~kg~m$^{-3}$ (solid line), $\rho_{\rm p} = 4 \times 10^{-11}$~kg~m$^{-3}$ (dotted line),  $\rho_{\rm p} = 3 \times 10^{-11}$~kg~m$^{-3}$ (dashed line), and $\rho_{\rm p} = 6 \times 10^{-11}$~kg~m$^{-3}$ (dot-dashed line). The shaded zone in panel $a)$ shows the region of phase speeds corresponding to external leaky slow modes.  \label{fig:flowfast}}
\end{figure*}

On the other hand, we see that the damping time of both slow and thermal modes is not affected by the presence of flow, as in \citet{carbonellflow}. Nevertheless, the attenuation of the fast kink mode in the present case is influenced by the flow. The larger the flow velocity, the more attenuated the parallel fast kink wave, whereas the opposite occurs for the anti-parallel solution. This behavior can be understood with the help of Figure~\ref{fig:flowfast}, which displays the phase speed of the parallel fast kink mode and its damping time for a wider range of the flow velocity and for different values of the thread density. We see that for a specific value of the flow velocity the parallel fast mode phase speed coincides with that of the external leaky slow modes, that is in the range $\left[ \Re \left( \tildectc \right) , \Re \left(  \lambdamodc \right) \right]$. Then, for such flow velocity, the parallel fast kink wave couples with the external slow modes by means of a ``weak'' coupling, according to the nomenclature from \citet{soler2}. This coupling causes a minimum in the damping time as is clearly seen in Figure~\ref{fig:flowfast}b. On the other hand, a similar argument can be adopted to explain the increase of the anti-parallel fast mode damping time with $U_{\rm p}$, since the phase speed of the anti-parallel wave moves away to the phase speed region of (anti-parallel) external slow modes, i.e. $\left[ -\Re \left(  \lambdamodc \right) , - \Re \left( \tildectc \right) \right]$, as the flow speed increases. For parallel waves, the flow velocity for which the coupling takes place depends on the thread density, in such a way that the smaller the density, the smaller the flow velocity. Hence, the minimum of the damping time moves to smaller values of the flow velocity for a small thread density. Nevertheless, this minimum takes place at a flow velocity of about 40~km~s$^{-1}$ for a representative thread density of $\rho_{\rm p} = 5 \times 10^{-11}$~km~s$^{-1}$, which is a velocity larger than those observed by at least a factor of two. Moreover, even the smallest damping time is several orders of magnitude larger than the lifetimes of prominence threads, meaning that the effect of the flow is not enough to obtain a reasonable and realistic attenuation for kink modes.

\clearpage
\clearpage

\subsubsection{Non-adiabatic sound speed}
\label{sec:soundspeed}

\begin{figure}[!tb]
\centering
\epsscale{1}
\plotone{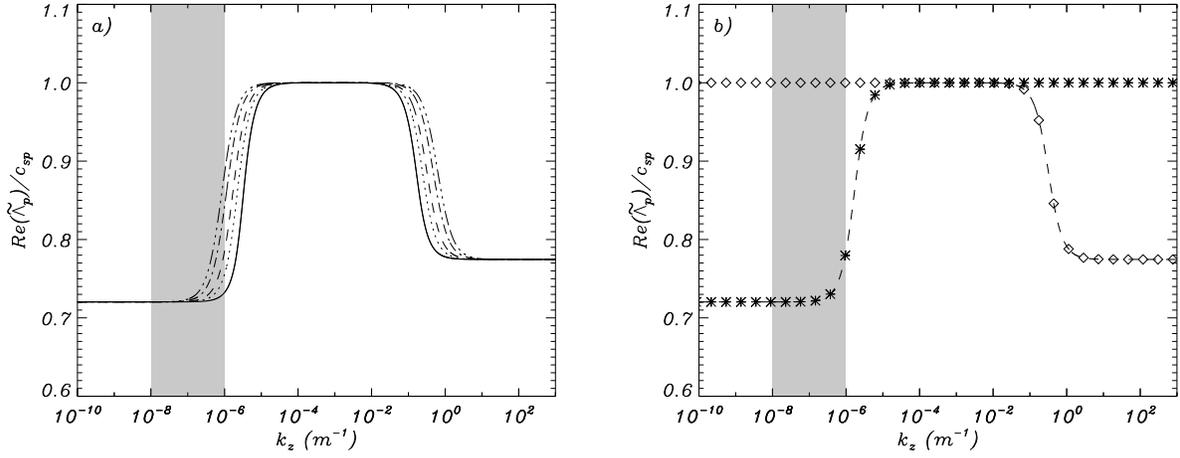}
\caption{Real part of the internal non-adiabatic sound speed (in units of the internal adiabatic sound speed) versus the longitudinal wavenumber. $a)$ Result for different flow velocities: $U_{\rm p} = 0\, {\rm km}\, {\rm s}^{-1}$ (solid line), $U_{\rm p} = 5\, {\rm km}\, {\rm s}^{-1}$ (dotted line), $U_{\rm p} = 10\, {\rm km}\, {\rm s}^{-1}$ (dashed line), $U_{\rm p} = 20\, {\rm km}\, {\rm s}^{-1}$ (dot-dashed line), and $U_{\rm p} = 30\, {\rm km}\, {\rm s}^{-1}$ (three dot-dashed line). $b)$ Result for $U_{\rm p} = 10\, {\rm km}\, {\rm s}^{-1}$ considering all non-adiabatic mechanisms (dashed line), neglecting prominence conduction (asterisks), and neglecting prominence radiation (diamonds). The shaded zone in both panels indicate the range of observed wavelengths. \label{fig:lambda}}
\end{figure}

Next, we study the influence of the steady mass flow on the value of the internal non-adiabatic sound speed, $\lambdamodp$ (eq.~[\ref{eq:lambda}]). Since $\Re \left( \lambdamodp \right)$ corresponds to the value of the flow velocity for which anti-parallel slow waves become backward waves, it is interesting to assess the behavior of $\Re \left( \lambdamodp \right)$ as a function of the wavenumber and the flow velocity. We perform a similar study as in \citet{carbonellflow} and compute the real part of $\lambdamodp$ as a function of $k_z$ for different values of the flow velocity (Fig.~\ref{fig:lambda}). The behavior of $\Re \left( \lambdamodp \right)$ with $k_z$ is the one explained by \citet{carbonellflow}, and presents three different plateaus. For large $k_z$ the dominant mechanism is prominence thermal conduction, and the non-adiabatic sound speed coincides with the isothermal value. For intermediate $k_z$ the non-adiabatic sound speed does not depend of non-ideal effects, and its value corresponds to the adiabatic sound speed. For small $k_z$, including the observed region of wavelengths, the non-adiabatic sound speed becomes
slightly smaller than the isothermal one, and prominence radiation is the governing mechanism. On the other hand, one can see that the effect of the flow is to widen the range of $k_z$ in which the non-adiabatic sound speed matches the adiabatic value: the larger the flow, the wider the range. For large flow velocities, the transition between the intermediate-$k_z$ (adiabatic) and the small-$k_z$ plateaus takes place within the observed wavelengths region.

\clearpage
\clearpage

\section{Conclusions}
\label{sec:conclusions}

In the present work, we have studied the combined effect of both non-adiabatic effects and a steady mass flow on the damping of oscillations supported by an individual, cylindrical, and homogeneous prominence thread. Our main conclusions are summarized next.
\begin{enumerate}
 \item In the absence of flow, slow modes are efficiently damped by non-adiabatic effects, while fast kink waves are in practice non-attenuated since their damping times are much larger than typical lifetimes of filament threads. 
\item The damping  by non-adiabatic mechanisms of transverse, kink oscillations is much less efficient in the present, cylindrical case than in the slab geometry.
\item The thermal wave is a non-propagating solution in the absence of flow and its attenuation by radiative losses is extremely quick.
\item The presence of flow breaks the symmetry between waves propagating parallel or anti-parallel to the flow. For a flow velocity larger than the internal non-adiabatic sound speed anti-parallel slow waves become backward waves.
\item In the presence of flow, the thermal mode behaves as a wave that propagates parallel to the flow and its motions are mainly polarized along the longitudinal direction. Nevertheless, this oscillatory behavior cannot be likely observed in practice due to its quick damping. 
\item The damping time of both slow and thermal waves is not affected by the flow. On the contrary, for realistic values of the flow velocity, the larger the flow, the larger the attenuation of parallel fast kink waves, whereas the contrary occurs for anti-parallel fast kink solutions. Nevertheless, this effect is not enough to obtain realistic damping times in the case of fast kink modes.
\end{enumerate}

In agreement with previous studies, the consideration of non-adiabatic mechanisms provides with damping times that are compatible with observations in the case of slow modes, which can be related to long-period oscillations. The main effect of the flow on these solutions is that only parallel propagation to the flow is allowed for strong enough flows. The negligible attenuation of fast kink modes in the absence of flow is slightly improved in the case of parallel waves when flow is present, the damping time being diminished by an order of magnitude for realistic flow velocities. However, the damping time is still several orders of magnitude larger than the lifetimes of filament threads and, therefore, neither non-adiabatic mechanisms nor mass flows provide with reasonable fast mode damping times applicable to prominences. For this reason, it is likely that another damping mechanism is responsible for a more efficient attenuation of transverse thread motions, resonant absorption being a good candidate which should be investigated. On the other hand, the investigation of the effect of flow, and in particular counter-streaming flows, on the damping of collective transverse oscillations of cylindrical multithread models is interesting in the light of the present results and could be the subject of a future research.

\acknowledgements
     The authors acknowledge the financial support received from the Spanish Minis\-terio de Ciencia y Tecnolog\'ia and the Conselleria d'Economia, Hisenda i Innovaci\'o under Grants No. AYA2006-07637 and PCTIB-2005GC3-03, respectively. R.~Soler thanks the Conselleria d'Economia, Hisenda i Innovaci\'o for a fellowship.

\appendix

\section{Expressions for the perturbations}
\label{app:pert}

Expressions for the perturbations as functions of the divergence of the perturbed velocity, $\Delta = \nabla \cdot {\mathit {\bf v}}_1$, and its derivative are given next,
\begin{equation}
 v_r = - \frac{\left( \Omegao^2 - k_z^2 \lambdamod^2 \right)}{\Omegao^2 m_0^2} \frac{\partial \Delta}{\partial r},
\end{equation}
\begin{equation}
 v_\varphi = - i \frac{\left( \Omegao^2 - k_z^2 \lambdamod^2 \right)}{\Omegao^2 m_0^2} \frac{n}{r} \Delta,
\end{equation}
\begin{equation}
 v_z = - i \frac{\lambdamod^2 k_z}{\Omegao^2} \Delta,
\end{equation}
\begin{equation}
 \rho_1 = i \frac{\rho_0}{\Omegao} \Delta,
\end{equation}
\begin{equation}
 p_1 = i \frac{\rho_0 \lambdamod^2}{\Omegao} \Delta,
\end{equation}
\begin{equation}
 T_1 = i \frac{T_0}{\Omegao} \left( \gamma \frac{\lambdamod^2}{\cs^2} - 1 \right) \Delta,
\end{equation}
\begin{equation}
B_{r} = - B_0 k_z \frac{\left( \Omegao^2 - k_z^2 \lambdamod^2 \right)}{\Omegao^3 m_0^2} \frac{\partial \Delta}{\partial r},
\end{equation}
\begin{equation}
B_\varphi = - i B_0 k_z \frac{\left( \Omegao^2 - k_z^2 \lambdamod^2 \right)}{\Omegao^3 m_0^2} \frac{n}{r}\Delta,
\end{equation}
\begin{equation}
 B_z = i B_0 \frac{\left( \Omegao^2 - k_z^2 \lambdamod^2 \right)}{\Omegao^3} \Delta,
\end{equation}
\begin{equation}
 p_{\rm m_1} = i \rho_0 \va^2 \frac{\left( \Omegao^2 - k_z^2 \lambdamod^2 \right)}{\Omegao^3} \Delta,
\end{equation}
\begin{equation}
 p_{\rm T_1} = p_1 + p_{\rm m_1} =  i \rho_0 \frac{\left( \Omegao^2 - k_z^2 \lambdamod^2 \right) \left( \Omegao^2 - k_z^2 \va^2 \right)}{\Omegao^3 m_0^2} \Delta.
\end{equation}

\section{Region of non-existence of the thermal mode}
\label{app:thermal}

\begin{figure}[!htb]
\centering
\epsscale{0.5}
\plotone{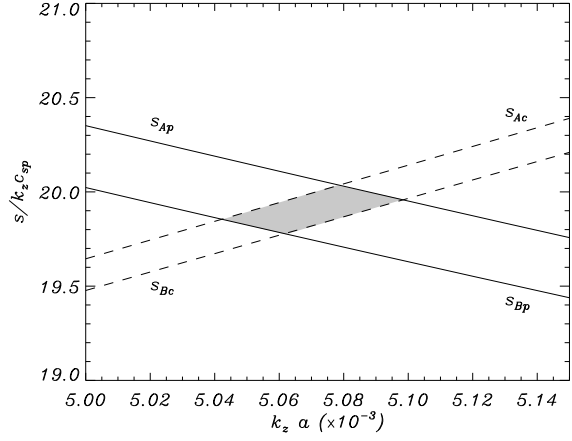}
\caption{Damping rate per wavenumber versus the wavenumber, both in dimensionless units. The shaded zone points the region where the thermal mode does not exist as an evanescent-like solution, i.e. the overlap of the regions where $m_{\rm p}^2 > 0$ (between solid lines) and $m_{\rm c}^2 > 0$ (between dashed lines) are both satisfied. \label{fig:thermrange}}
\end{figure}

Following a similar argument as in \citet{vanderlinden}, let us consider that the thermal mode frequency in the absence of flow is given by $\omega = i s$, with $s$ a real quantity. So, from equation~(\ref{eq:m0}) one obtains  
\begin{equation}
 m_0^2 = - \frac{\left( s^2 + k_z^2 \va^2 \right) \left( s^2 + k_z^2 \tilde{\Lambda}^2_0 \right)}
{\left( \va^2 + \tilde{\Lambda}^2_0 \right) \left( s^2 + k_z^2 \nct^2 \right)} = - \left( s^2 + k_z^2 \va^2 \right) \frac{\mathcal{A}}{\mathcal{B}}. \label{eq:m0thermal}
\end{equation}
Quantities $\mathcal{A}$ and $\mathcal{B}$ are the following third-order polynomials in $s$, 
\begin{eqnarray}
\mathcal{A} &=& s^3 - \mathcal{N}_2 s^2 + k_z^2 \cs^2 s - k_z^2 \cs^2 \mathcal{N}_2 , \\
 \mathcal{B} &=& s^3 - \mathcal{N}_3 s^2 + k_z^2 \ct^2 s -k_z^2 \ct^2 \mathcal{N}_1,
\end{eqnarray}
with
\begin{eqnarray}
 \mathcal{N}_1 &=& \frac{ \left( \gamma -1 \right) }{\gamma} \left( \frac{T_0}{p_0} \kappa_\parallel k_z^2 + \omega_T - \omega_\rho \right), \\
\mathcal{N}_2 &=&  \left( \gamma -1 \right) \left( \frac{T_0}{p_0} \kappa_\parallel k_z^2 + \omega_T \right), \\
\mathcal{N}_3 &=& \frac{\mathcal{N}_2 \va^2 + \mathcal{N}_1 \cs^2}{\va^2 + \cs^2}.
\end{eqnarray}
The condition $m_{\rm p}^2 > 0$ implies that ${\rm sign} \left( \mathcal{A}_{\rm p} \right) \neq {\rm sign} \left( \mathcal{B}_{\rm p} \right)$.  The solutions of $\mathcal{A}=0$ are a pair of complex conjugate roots and a real root, while the same stands for the roots of $\mathcal{B}=0$. Then, the condition $m_{\rm p}^2 > 0$ is only verified in the region between the real roots of $\mathcal{A}_{\rm p}=0$ and $\mathcal{B}_{\rm p}=0$, namely $s_\mathcal{A_{\rm p}}$ and $s_\mathcal{B_{\rm p}}$, respectively, which are very close to each other. On the other hand, the external evanescent requirement ($m_{\rm c}^2 < 0$) is verified outside the region between the real solutions of $\mathcal{A}_{\rm c}=0$ and $\mathcal{B}_{\rm c}=0$, namely $s_\mathcal{A_{\rm c}}$ and $s_\mathcal{B_{\rm c}}$ respectively. By computing these real roots (Fig.~\ref{fig:thermrange}), one obtains that both regions do not overlap except for $5.04 \times 10^{-3} \lesssim k_z a \lesssim 5.10 \times 10^{-3}$. Thus, outside this extremely narrow overlapping region, the thermal mode exists with a damping rate in the range $s_\mathcal{B_{\rm p}} < s < s_\mathcal{A_{\rm p}}$ where both conditions $m_{\rm p}^2 > 0$ and $m_{\rm c}^2 < 0$ are satisfied. 

\clearpage

\end{document}